# Reservoir-Engineered Squeezed Lasing through the Parametric Coupling

Yuhang Tian,[1,*] Yajun Wang,[1,2,*,†] Weijie Wang,[1] Xiaocong Sun,[3] Yuhang Li,[4]
Shaoping Shi,[1,2] Long Tian,[1,2] and Yaohui Zheng[1,2,‡]

[1]*State Key Laboratory of Quantum Optics Technologies and Devices, Institute of Opto-Electronics, Shanxi University, Taiyuan 030006, China*
[2]*Collaborative Innovation Center of Extreme Optics, Shanxi University, Taiyuan 030006, China*
[3]*School of Applied Science, Taiyuan University of Science and Technology, Taiyuan 030024, China*
[4]*Department of Precision Instrument, Tsinghua University, Beijing 100084, China*

We report the first experimental demonstration of squeezed lasing in a reservoir-engineered optical parametric oscillator (OPO). The OPO provides a basis of squeezed states and parametric amplification in lasing emission, whose vacuum reservoir is coupled to a squeezed vacuum generated by a second OPO. With a precisely controlled squeezing angle and strong squeezing injection, the parametric interaction in the first OPO is exponentially enhanced. It successfully circumvents the decoherence in the system, and eliminates the undesired noise of spontaneous photon emission in the OPO. As a result, the amplified parametric process simultaneously reserves the coherence and quantum properties in the first OPO, and yields a −6.1 dB squeezed laser in optical domain with a narrow linewidth and high brightness. Our work sheds light on potential applications of squeezed lasing in quantum metrology and quantum optics.

A squeezed laser retains both lasing coherence and photon correlations of squeezed states. It has potential applications that go beyond standard lasers or squeezed vacuum states. To improve the performance of coherent coupling between squeezed lasers and atomic transitions, squeezed states should have a narrow linewidth to drive superior applications of atomic quantum metrology [1,2], quantum memory, etc. [3,4]. A squeezed laser with lower noise and long coherence time also offers a possibility to improve the detection sensitivity, resolution, and distance in coherent detection scenarios, etc. [5–10], while it is also advanced for quantum key distribution maintaining a laserlike coherent amplitude [11,12]. In optical interferometry, a squeezed laser is free of squeezed vacuum state injection from the dark port of the interferometer [7,13,14], which is expected to simplify the implementation of precision measurement. These broad prospective applications generate great interest in researching squeezed lasing [15,16].

Conventionally, the subthreshold optical parametric oscillator (OPO) belongs to the most successful approaches for squeezed state generation [17–26], in which spontaneous parametric coupling induces a coherence deterioration [27,28]. When above threshold, the stimulated parametric coupling induces lasing radiation with excellent coherence [27,28]. However, two-photon damping and thermal noise emerge to induce an undesired noise [16,29–33], along with a phase transition to turn the squeezed state into a mixture of two coherent states [34]. In this regard, the coherence and quantum properties herein are mutually incompatible. Additionally, the output frequency spectrum of an OPO always has a frequency comb property with a frequency separation of its free spectral range [19,35–37], which forms a multimode optical field output with multiple frequency components to cause a mode competition in lasing emission. Hence, in order to prepare a squeezed laser with an OPO, we should eliminate the undesired noise under single mode operation. A detuned parametric driven cavity with an external squeezed vacuum reservoir and an intracavity atom gain medium is proposed to prepare a squeezed laser [16]. Profiting from the squeezed vacuum reservoir, the undesired noise is removed, and the photon-atom interaction is simultaneously enhanced by a factor of $\cosh 2r$ ($r$ is the squeezing parameter) [16,29,30,38,39]. Finally, a single squeezed laser mode corresponding to the transition of the atom energy levels can be directly generated. However, the proposal is confronted with a great challenge in experimental implementation. Strikingly, an open quantum system with assistance of squeezed vacuum not only can be parametrically driven to generate a desired quantum state with undesired noise removed, but also can enhance the interaction in the system to become a strong coupling one and preserve longer coherence time [15,16,29,30,38–40]. The regime provides a feasible route for squeezed lasing generation.

[*]These authors contributed equally to this work.
[†]Contact author: YJWangsxu@sxu.edu.cn
[‡]Contact author: yhzheng@sxu.edu.cn

Here, we report on the first experimental demonstration of squeezed lasing in an optical domain, resorting to a low-loss OPO engineered by a broadband squeezed vacuum. In contrast to the above threshold OPO, the reservoir-engineered OPO enables to enhance the parametric interaction and reduce the noise coupling with environment simultaneously, which drives the spontaneous parametric coupling into a stimulated parametric one, and reserves the quantum property of the output field. Because of the difference in cavity length design of the two OPOs for squeezed vacuum and squeezed laser generation, only the single cavity mode at the carrier frequency can be amplified to form a lasing. As its origin from a subthreshold OPO, the process simultaneously preserves the squeezing and laser-like coherence properties. This experiment points toward the potential applications of squeezed lasing in quantum metrology and quantum optics.

*Optical parametric coupling with engineered reservoir*—Let us consider a $\chi^{(2)}$ parametric coupling process with subthreshold OPO for single mode squeezed state generation in Figs. 1(c) and 1(d), which is validly demonstrated in Sec. 3 of Supplemental Material [41]. A pump photon with a carrier frequency of $\omega_p$ spontaneously down-converts into a signal photon $\omega_s$ and an idle photon $\omega_i$ [Fig. 1(c)], with coherence deterioration [Fig. 1(d)] [27,28]. When the OPO operates above threshold, the stimulated down-conversion dominates the interaction with perfect coherence [Figs. 1(a) and 1(b)] [27,28]. However, due to the stimulated intense signal and idler fields, two-photon damping [Fig. 1(a)] occurs to cause a phase transition at threshold, accompanied by spontaneous symmetry breaking of the discrete phase above threshold [34]. It turns the squeezed vacuum state into a mixture of two coherent states [Fig. 1(b)] [16,34]. The influence of the damping can be also treated as an undesired noise coupling to submerge the squeezed property, as a consequence of the vacuum input [16,29,30,38,39]. Hence, the conventional OPO cannot be directly applied to generate a squeezed lasing.

Recently, a parametrically driven OPO with squeezed vacuum reservoir had been theoretically proposed to eliminate the two-photon damping induced undesired noise, and exponentially enhance the light-matter interactions at the same time in the atom [16,29,39] and optomechanical resonators systems [30,38]. Inspired by these ideologies, we propose a squeezed vacuum reservoir-engineered OPO to prepare a squeezed laser [Figs. 1(e) and 1(f)]. Differing from Ref. [16], the parametric interaction in our system is reserved without cavity detuning. With assistance of squeezed vacuum reservoir, the intracavity parametric coupling is exponentially enhanced with undesired noise removed. Then, the Hamiltonian of the system can be written as $\hat{H} = \Delta_c \hat{a}^+\hat{a} + g(e^{-i\theta_P}\hat{a}^2 + \text{H.c.})/2$, where $\theta_P$ is the phase of the pump field, $\Delta_c$ is the detuning of the cavity, and $\hat{a}$ represents the annihilation operator. By changing the basis with squeezing unitary transformation $\hat{S}(\xi) = \exp[1/2(\xi^*\hat{a}^2 - \xi\hat{a}^{+2})]$, where $\xi = re^{-i\theta}$, we substitute $\hat{a} \to \hat{a}_s \cosh r - \hat{a}_s^+ e^{-i\theta} \sinh r$, in which $\hat{a}_s$ is the annihilation operator in the new squeezed basis, and $\theta$ is the squeezing angle in squeezing unitary transformation. Then, the Hamiltonian in the new basis can be deduced as [16,29,41]

$$\begin{aligned}\hat{H}_s =& \left\{\hat{a}_s^2\left[\frac{g}{2}(e^{i\theta_P}\cosh^2 r + \sinh^2 r e^{-i(\theta_P - 2\theta)})\right.\right.\\ &\left.\left. - \Delta_c e^{i\theta} \cosh r \sinh r\right] + \text{H.c.}\right\}\\ &+ \hat{a}_s^+\hat{a}_s\left[-g\cosh r \sinh r\left(e^{i(\theta_P - \theta)} + e^{-i(\theta_P - \theta)}\right)\right.\\ &\left. + (\cosh^2 r + \sinh^2 r)\Delta_c\right]\\ &- \frac{g}{2}\cosh r \sinh r(e^{i(\theta_P - \theta)} + e^{-i(\theta_P - \theta)}) + \Delta_c \sinh^2 r,\end{aligned} \quad (1)$$

where $g$ is the parametric coupling strength, and $r$ is the squeezing parameter. Apparently, this Hamiltonian offers a possibility to modify the parametric interaction, and engineer the squeezed vacuum reservoir by controlling the squeezing angle $\theta$ and parameter $r$. We set $\theta_P - \theta = \pi$; the interaction term of $\hat{H}_s$ reduces to $\hat{a}_s^2[-(g_s/2) - \Delta_c \cosh r \sinh r] + \text{H.c.}$. With $\Delta_c = 0$, the resulting parametric coupling strength $g_s$ is exponentially enhanced by a factor of $\cosh 2r$, i.e., $g_s = g\cosh 2r$ [16,29,30,38–41].

Taking a phase squeezed vacuum reservoir coupled to an amplitude squeezed basis as an example, i.e., $\theta_P = \pi$ and $\theta = 0$, we substitute the Hamiltonian of Eq. (1) into the conventional motion equation of the OPO, the quadrature variance of amplitude $\hat{X}_{\text{out}}$ of the squeezed laser can be derived as [41]

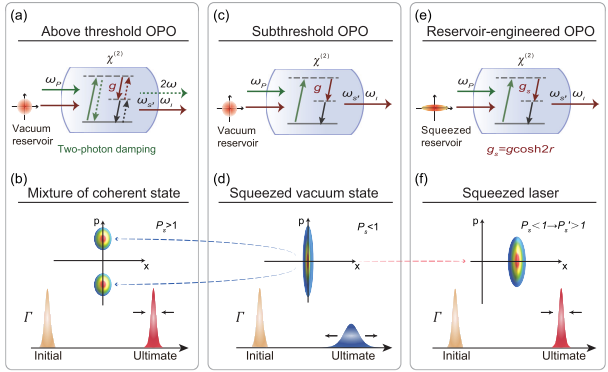

FIG. 1. Physical mechanism of the parametric coupling with vacuum or squeezed vacuum reservoir. (a) Parametric down-conversion (PDC) in an above threshold OPO. (b) Mixture of two coherent states with coherence preservation and quantum characteristic deterioration. (c) PDC in a subthreshold OPO. (d) Squeezed vacuum state with quantum characteristic preservation and coherence deterioration. (e) PDC in a subthreshold OPO with squeezing vacuum reservoir. (f) Squeezed lasing with coherence and quantum characteristic preservation.

$$\delta^2 \hat{x}_{\text{out}} = \left(\frac{\kappa^2}{2M} + \sqrt{1-\kappa}\right)^2 e^{2r}$$
$$- \left[\frac{\kappa}{M}(-i\omega + A - B)\right]^2 e^{-2r}, \qquad (2)$$

where $A = -(i\Delta_c/\hbar)\cosh 2r - ig\sinh 2r$, $B = (i\Delta_c/\hbar)\sinh 2r + ig_s$, $M = (i\omega - A + \sqrt{\kappa})(-i\omega - A^* + \sqrt{\kappa})$, $\kappa$ is the cavity damping constant, and $\omega$ is the sideband frequency of the carrier. The detailed derivation is shown in Supplemental Material [41].

*Experimental realization of squeezed laser*—Figure 2 illustrates the schematic diagram of our experimental setup. A pump laser with the wavelength of 532 nm is prepared by a second harmonic generator (SHG). Its output is divided into two beams to provide the pump sources of the two OPOs. The OPOs are both semimonolithic resonators, which consist of a concave mirror and a PPKTP crystal. When they work independently, more than $-12/-10$ dB amplitude squeezing had been observed [42,45]. It indicates a high squeezing parameter and low intracavity loss is available, which provides the basic conditions for squeezed lasing generation. OPO1 with a linewidth of 68 MHz and a free spectral range of 3.33 GHz is applied for squeezed vacuum preparation, and OPO2 plays a crucial role for squeezed basis supply and parametric amplification, which linewidth is 97 MHz and free spectral range is 3.76 GHz. In lasing emission, OPO1 operates at parametric amplification status with phase quadrature squeezing, while OPO2 is locked at opposite phase versus OPO1 to generate an amplitude squeezed light. The small length difference of the two OPOs ensures a stable single mode operation under squeezing enhanced regime, which ensures only one cavity mode of OPO2 in the center of the PPKTP spectral acceptance bandwidth is exponentially amplified. The detailed parameters of the OPOs and laser properties can be found in Supplemental Material [41].

Of the squeezed vacuum existing from OPO1, 99% transmits through a Faraday isolator (FI), and is mode matched to OPO2 with an efficiency of 99% to excite the exponentially amplified parametric coupling. Of the squeezing signal, 1% is coupled with a coherent control beam (~100 μW) [46], which passes through a two-stage acousto-optic modulator (AOM) to produce a frequency shift beam of 20 MHz, and is demodulated at 40 MHz to create an error signal to lock the relative phase between the control and squeezing beams to 0. The coupled beam acts as an auxiliary signal to control the relative phase between the squeezed laser and pump field to $\pi$. The frequency shifted beam has a linewidth of 300 kHz, i.e., $20 \pm 0.15$ MHz, which has no risk for the sideband noise variance evaluation at 18 MHz. Owing to the superiority and advancement of our OPOs and control technologies [42,45,47,48], the squeezed lasing can be continuously radiated from OPO2. The generated squeezed laser exits from the output coupler of OPO2, and is reflected by the input port of FI. Its noise variance of the amplitude quadrature at sideband frequency and laser linewidth are measured by the self-homodyne detection [49] and $\beta$-separation line algorithm [41,50] methods, respectively. In linewidth characterization, an analysis cavity is utilized to calibrate the frequency noise spectrum of the laser [33,43,44,49].

*Results and discussion*—The experimental arrangements supply a squeezed vacuum reservoir and guarantee an amplitude squeezed lasing, in which only the generated photons with carrier frequency are exponentially amplified. To comprehensively evaluate the squeezed lasing performance, we optimize the pump ratio and squeezing parameters of the OPOs to excavate the transformation law for the optimum laser power, squeezing level, and coherence, as shown in Figs. 3 and 4.

Squeezed laser power versus pump power $P$: A $-10.4$ dB squeezed light of phase quadrature generated from OPO1 is coherently coupled to OPO2. Propagation loss sets the fundamental limit to the possible enhancement of parametric amplification in squeezed lasing. The expected losses for components between the two OPOs are 5%, measured by the powers of the signal beam at the output port of OPO1 and input port of OPO2. The remainder squeezing coupled to OPO2 is reduced to $-8.6$ dB, which supplies a 3.7 times enhancement for parametric gain. The result can be also confirmed by the changes of threshold power $P_{\text{th}}$ of OPO2, which decreases from $P_{\text{th}} = 45$ mW to $P'_{\text{th}} = 11.4$ mW (more details are provided in Supplemental Material). With constant squeezed vacuum injection, the output power of the squeezed laser against the pump ratio $P'_s = P/P'_{\text{th}}$, normalized to the new threshold power of OPO2, is measured by a power meter, as shown in Fig. 3(a). The laser output power increases from 0.5 to 2.6 mW as $P'_s$ changes from 1 to 1.75, corresponding to the pump powers from 11.4 to 20 mW. It is markedly below the physical threshold power with vacuum reservoir. To increase $P'_s$ further, the squeezed lasing becomes unstable. We speculate two reasons contribute to this effect, one is the multiple cavity-mode overlapping shown in Fig. S3 of Supplemental Material [41], which results in multimode competition; the other one is more undesired noise production due to higher pump strength, which causes the relative phase $\theta_P$ drifting from $\pi$ and unlocked.

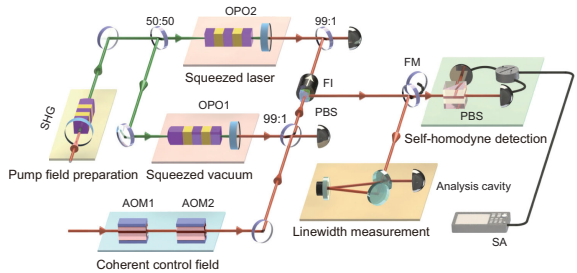

FIG. 2. Schematic of the experimental setup for squeezed laser generation and evaluation. HR, high reflectivity; FI, Faraday isolator; PBS, polarization beam splitter; SHG, second harmonic generator; OPO1,2, optical parametric oscillator; AOM, acousto-optic modulator; BS, beam splitter; FM, folding mirror; SA, electrical spectrum analyzer.

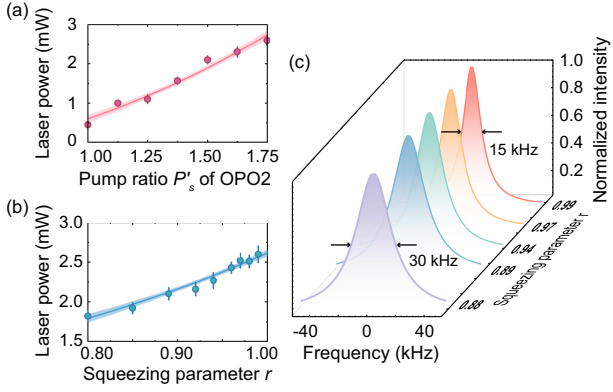

FIG. 3. (a) Squeezed laser power versus pump ratio $P'_s$ of OPO2 with reservoir engineered by a $-10.4$ dB squeezed state of phase quadrature. (b) Squeezed laser power at the pump ratio of $P'_s = 1.75$ versus the squeezing parametric $r$. (c) Variation of laser linewidth with the squeezing parametric $r$ measured by $\beta$-separation line algorithm method. The data (dots) are fitted with analytical expressions (solid lines) adopted from (S2.9) and (S2.10) in Supplemental Material [41]. The results show that the squeezed laser power increases linearly with pump power (red curve), and increases with the squeezing parameter $r$ by a factor of $\cosh 2r$ (blue curve). The data is fitted with 68% confidence bands (shaded area) and the error bars stand for 1 s.d.

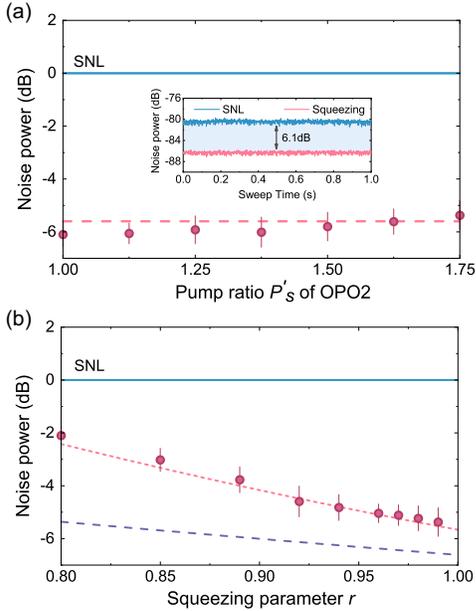

FIG. 4. Noise power of the squeezed laser versus pump ratio $P'_s$ (a) and injected squeezing parameter $r$ (b). Red solid dot points, experimental data; red dashed lines, theoretical fitting with undesired noise; blue dashed line, theoretical result without undesired noise. The theoretical curves are followed by equations (S2.13) and (S2.14) in Supplemental Material. Blue solid line represents the shot noise limit (SNL) level. All the data are measured at the sideband frequency of 18 MHz. RBW, 300 kHz; VBW, 200 Hz. The error bars stand for 1 s.d.

Squeezed laser power versus squeezing parameter $r$: With $P'_s = 1.75$, we also investigate the laser power versus the squeezing parameter $r$, which is continuously controlled by the pump ratio of OPO1. The results are shown in Fig. 3(b), which indicate the power is monotonically amplified to a maximum value of 2.6 mW along with $r$ increased from 0.8 to 0.99. The observed power increment represents the direct experimental manifestation of lasing behavior driven by a squeezed vacuum to stimulate a macroscopic photonic occupation with squeezing property. Additionally, the phase squeezed light from OPO1 has a bright amplitude, which power is raised with $r$ increasing, and is seeded into OPO2 [41]. But we have experimentally demonstrated that this weak mean field has negligible influence to the squeezed laser power due to its opposite phase to that of the squeezed laser, and only the photons generated in the cavity are amplified [41]. Therefore, the power of the bright squeezed light is directly deducted from the results in Figs. 3(a) and 3(b). A detailed analysis can be found in Supplemental Material [41].

Laser linewidth evolution: Laser linewidth $\Gamma$ is connected with the coherence of a laser, which is one of the key features of the squeezed lasing distinguishing with the squeezed vacuum state. We also synchronously evaluate the linewidth enhanced by squeezed vacuum under the identical experimental circumstances of (b), which is characterized by a $\beta$-separation line algorithm method [41]. In the first step, we calibrate the accuracy of the $\beta$-separation line algorithm method with the standard delayed self-heterodyne one [51]. Taking the initial 1064 nm laser source used in the experiment as an example, the measured linewidths are 13 kHz (delayed self-heterodyne) and 15 kHz ($\beta$-separation line algorithm), respectively. The tiny difference between the two technologies certifies that $\beta$-separation line method is reliable to evaluate the linewidth of squeezed laser. Next, the improvement of the linewidth $\Gamma$ with squeezing enhancement is characterized as shown in Fig. 3(c). When $r$ changes from 0.8 to 0.99, the linewidth of the squeezed laser reduces from 30 to 15 kHz. It is apparent that the linewidth gradually approaches that of the original laser. The linewidth preservation also gives the credit to strong coupling mechanism with assistance of squeezed vacuum reservoir in OPO2, which stimulates a coherent parametric interaction between the photons and dipoles to circumvent the optical decaying in the system.

Squeezing property preservation: The final essential feature of a squeezed laser is the quantum correlation. Figure 4 shows the squeezing degree of the squeezed laser at the sideband frequency of 18 MHz versus the pump ratio

$P'_s$ of OPO2 and the injected squeezing parameter $r$. Figure 4(a) depicts a maximum squeezing level with the same experimental conditions of Fig. 3(a). With the injected squeezing parametric $r = 0.99$, the squeezing degrees are basically close to −6 dB, slowly decreasing with the increasing of pump ratio $P'_s$, which is mainly attributed to more undesired noise coupling with stronger pump strength. With $P'_s = 1$, the optimal squeezing is observed to be −6.1 dB, which is mainly limited by the propagation loss between OPO1, OPO2, and the self-homodyne detection unit; more details can be found in Supplemental Material [41]. For $P'_s = 1.75$, we change the injected squeezing parametric $r$ from 0.8 to 0.99, the noise power of the sideband frequency evolves from −3.2 to −5.4 dB, shown in Fig. 4(b). By comparing the difference between the theoretical results with and without undesired noise [the red and blue dashed lines in Fig. 4(b)], we can infer that the residual undesired noise exists in the system, which decreases the squeezing level of the squeezed laser. Distinctly, the higher the injected squeezing parametric $r$, the stronger the generated squeezing is, because of more undesired noise is suppressed with stronger squeezing input. The deviation between the theoretical (red dashed line) and experimental results maybe mainly come from the uncertainty of the threshold power calibration, and nonlinear noise coupling from the pump beam. Therefore, it can be predicted that more than 6.1 dB squeezing is available, e.g., in our state-of-the-art OPO [18,52], quantum noise in the sideband frequency can be squeezed more than −13.1 dB [41], but only if we further reduce the propagation losses between the two OPOs, and increase their squeezing level.

We perform the entire measurement several times on different days to show reproducibility of the regime. Our experimental results are in good agreement with the theoretical analysis, and demonstrate an efficient protocol for squeezed lasing emission with a reservoir-engineered OPO regime. The experimental scheme can be utilized to circumvent the interaction with environment, and preserve both the coherence and quantum properties of the system.

*Conclusion and outlook*—We have shown how to engineer the vacuum reservoir of a OPO with assistance of a squeezed vacuum state that emits laser into a squeezed state in a parametric coupling scenario. Our scheme relies on two squeezed states of opposite phases to exponentially amplify the parametric interaction: for one thing it turns the spontaneous photon emission into a stimulated one to perform a laser radiation, while for the other it allows engineering the vacuum reservoir with squeezed state injection to drive the lasing action into a squeezing operation before phase transition. Therefore, the coherence of a laser and quantum property of a squeezed state can be simultaneously preserved in one laser beam. For the first time, a 15 kHz narrow linewidth, −6.1 dB squeezed laser with an output power of 2.6 mW is experimentally generated in optical domain. Our demonstration provides a promising route for implementing squeezed lasing with current state-of-the-art squeezed state preparation technology, while further reducing the propagation loss, and more stable phase-locking improvement would produce a stronger squeezing more than 10 dB and more robust operated nonclassical lasing. Our work opens a possibility to simultaneously preserve the coherence and quantum properties of a squeezed lasing regime, and may find applications in quantum sensing and metrology.

*Acknowledgments*—We acknowledge fruitful discussions and insights with Heng Shen. We acknowledge financial support from the National Natural Science Foundation of China (NSFC) (Grants No. 62225504, No. 62027821, No. U22A6003, No. 62375162, No. 12304403), National Key Research and Development Program of China (No. 2024YFF0726401).

# Supporting Information for "Reservoir-engineered squeezed lasing through the parametric coupling"

Yuhang Tian[1,*], Yajun Wang[1,2,*,†], Weijie Wang[1], Xiaocong Sun[3], Yuhang Li[4], Shaoping Shi[1,2], Long Tian[1,2], Yaohui Zheng[1,2,‡]

[1]State Key Laboratory of Quantum Optics Technologies and Devices, Institute of Opto-Electronics, Shanxi University, Taiyuan 030006, China

[2]Collaborative Innovation Center of Extreme Optics, Shanxi University, Taiyuan, Shanxi 030006, China

[3]School of Applied Science, Taiyuan University of Science and Technology, Taiyuan 030024, China

[4]Department of Precision Instrument, Tsinghua University, Beijing 100084, China

This is the supplementary material for the paper: "Reservoir-engineered squeezed lasing through the parametric coupling". The purpose of this material is to present the details of squeezed laser: S1. Experimental implementation; S2. Basic theoretical model; S3. Laser property of the squeezed lasing.

## S1. Experimental implementation

### (i) Experimental configuration of the squeezed lasing preparation

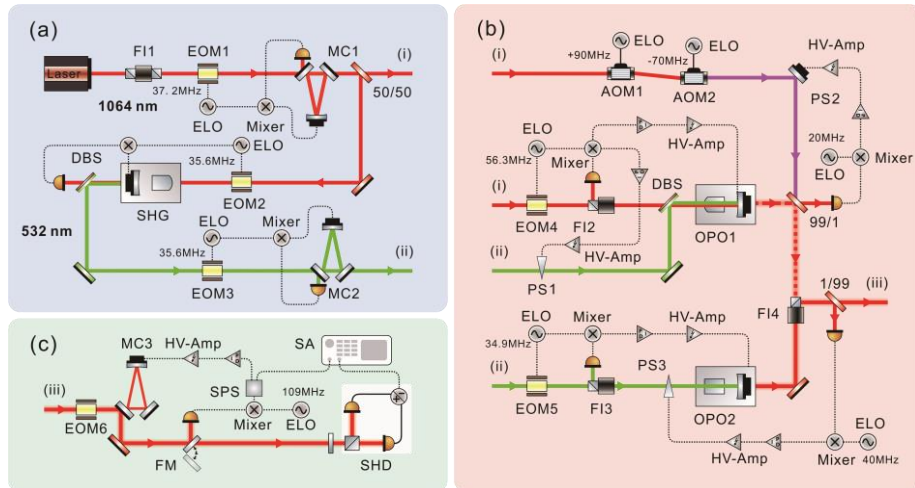

Fig. S1 Schematic of the experimental setup for squeezed laser generation and detection. EOM, electro-optical modulator; ELO, electronic local oscillator; PID, Proportional-Intergral-Differential servo; HV-Amp, High voltage amplifier; AOM, acousto-optical modulator; DBS, dichroic beam splitter; FM, flip mirror; FI, faraday isolator; MC, mode cleaner; SHG, second harmonic generator; PS, phase shifter; SPS, signal power splitter; OPO1,2, optical parametric oscillator;

SHD, self-homodyne detection; SA, spectrum analyzer.

Fig. S1 shows the experimental setup for squeezed lasing in detail. The fundamental wave laser source is a continuous-wave single-frequency solid-state laser with an output power of 2 W at 1064 nm. Firstly, the laser passes through a faraday isolator (FI1) to prevent optical feedback effect from the downstream optical elements. Secondly, a mode cleaner (MC1) is used to stabilize the laser power and frequency noises above its linewidth, meanwhile improves the laser spatial transverse mode [S1]. Then, it is divided into two parts: one is injected into the second harmonic generator (SHG) to produce a 532 nm laser, which serves as the pump fields for the two optical parametric oscillators (OPOs); the other serves as the sensing beam of Pound-Drever-Hall (PDH) technique for the relative phases and cavity lengths locking of the two OPOs [S2]. The two OPOs are both semi-monolithic cavities, consisting of a piezo-actuated concave mirror and a periodically poled $KTiOPO_4$ (PPKTP) crystal with the dimensions of 10 mm*2 mm*1 mm. OPO1 is a single-resonant cavity for the fundamental wave. One of the PPKTP crystal end face of OPO1 is polished to convexity with a 12 mm curvature radius, and coated with high reflectivity (HR) at 1064 nm (>99.9%) and high transmission (HT) at 532 nm (>98%). A see beam with 2.5 mW is injected from this face applied for cavity length and relative phase locking. The plane face is anti-reflection (AR) coated for both wavelengths (<0.2%). The concave mirror has a radius of curvature of 30 mm with a transmissivity of 12% for 1064 nm and HR for 532 nm. OPO2 is a doubly-resonant one, in which the fundamental and second harmonic waves are co-resonate. The end faces of the PPKTP in OPO2 are plane. One is serves as a cavity mirror with HR 1064 nm (>99.9%) and 532 nm (>98%) coating film, and the other is AR coated for both wavelengths (<0.2%). The concave mirror with a curvature radius of 25 mm serves as the output coupler, and is coated with reflectivity of 85% at 1064 nm and 97.5% at 532 nm, respectively.

OPO1 is applied to produce a phase quadrature squeezed light of -10.4 dB, in which the relative phase between the pump and signal waves is locked to 0. In OPO2, in order to realize a continuous squeezed lasing output, a coherent control technique is employed to lock the relative phase between the signal and the pump waves to $\pi$ [S3]. The coherent control wave with a 20 MHz single sideband is prepared by two acousto-optic modulators (AOMs) with modulation frequencies of +90 MHz and -70 MHz, respectively. Then, it is coherently coupled with the squeezed light exiting from OPO1 at the 99/1 beamsplitter with 0 phase. 99% of this light is injected into OPO2, and serves as the sensing wave for the locking of $\pi$ relative phase between the signal and pump waves of OPO2. Finally, an amplitude squeezed lasing can be continuously produced. We can measure the noise variance of the squeezed quadrature with self-homodyne detection (SHD). The laser linewidth is calibrated by $\beta$-separation line method.

### (ii) High degree squeezing generation with Low-loss OPO

Table. S1 Losses budget and detailed parameters of the two OPOs.

| Source of loss | OPO1 (%) | OPO2 (%) |
|---|---|---|
| Round-trip loss $l$ | 0.15 | 0.4 |
| OPO escape efficiency $\eta_{esc}$ | 98.7 | 97.4 |
| Propagation efficiency $\eta_{pro}$ | 99.2 | 98.7 |
| Interference efficiency $\eta_{vis}$ | 99.8 | 99.6 |
| Quantum efficiency $\eta_{qe}$ | 99 | 99 |
| **Parameter** | **OPO1** | **OPO2** |
| Total efficiency $\eta_{tot}$ | 97 % | 94 % |
| Total phase fluctuation $\theta_{tot}$ | 30 mrad | 13 mrad |
| Threshold power $P_{th}$ | 90 mW | 45 mW |
| Analysis frequency $f$ | 18 MHz | 18 MHz |
| Linewidth $\gamma$ | 68 MHz | 97 MHz |
| Relative phase | 0 | $\pi$ |
| Squeezing degree | -10.4 dB (phase squeezing) | -10.0 dB (amplitude squeezing) |

To generate a squeezed laser, the OPOs must be decoupled from the environment. It is to say that the decays and losses in the cavity should be as low as possible. In general, the squeezed quadrature noise variance related to the decays and losses is expressed as [S4]

$$V_- = \left[1 - \frac{4*\eta_{tot}*\sqrt{P/P_{th}}}{\left(1+\sqrt{P/P_{th}}\right)^2 + 4*(f/\gamma)^2}\right]\cos^2\theta_{tot} + \left[1 + \frac{4*\eta_{tot}*\sqrt{P/P_{th}}}{\left(1-\sqrt{P/P_{th}}\right)^2 + 4*(f/\gamma)^2}\right]\sin^2\theta_{tot} \quad (S1.1)$$

where $P_{th}$, $f$, $\gamma$, $\eta_{tot}$ and $\theta_{tot}$ are the threshold power, analysis frequency, linewidth, total efficiency and phase fluctuation, respectively. $\eta_{tot} = \eta_{esc}\eta_{pro}\eta_{vis}\eta_{qe}$, $\eta_{esc} = T/(T+l)$, $T$ is the transmissivity of the output coupler and $l$ is the round-trip loss. The corresponding squeezing degree and parameter are read as $R_- = -10\,log_{10}\,V_-$ and $r = -\frac{1}{2}ln\,V_-$, respectively.

In the experiment, the directly measured squeezing degree of OPO1 is -10.4 dB at the pump power of 70 mW, while the corresponding squeezing parameter is $r$=1.2. The results demonstrate a low loss condition for squeezing laser generation. We also budget the losses of the squeezing preparation, propagation and detection process in Table S1 and S2. It can be found that the squeezing experiences 5% loss when transmits to OPO2. Then, the squeezing degree is reduced to -8.6 dB, corresponding to a squeezing parameter $r$ of 0.99. Therefore, the efficient squeezing degree of the generated squeezed laser in OPO2 is -6.6 dB, which is predicted by Eq. (S2.13).

Table. S2 Losses budget between the OPO1 and OPO2, and OPO2 to SHD.

| PBS | IO | 99:1 | Mode mismatch | Loss from OPO1 to OPO2 | Detection loss of SHD | Loss from OPO2 to SHD |
|---|---|---|---|---|---|---|
| 2% | 1% | 1% | 1% | 5% | 3% | 7% |

## S2. Basic theoretical model

The Hamiltonian of a parametrically-driven OPO2 without considering undesired noise can be described by the electromagnetic field annihilation operator $\hat{a}$ and creation operator $\hat{a}^+$, and is expressed as

$$\hat{H} = \Delta_c \hat{a}^+\hat{a} + \frac{1}{2}\hbar g e^{i\theta_p}\hat{a}^2 + \frac{1}{2}\hbar g e^{-i\theta_p}(\hat{a}^+)^2 \quad (S2.1)$$

where $\Delta_c$ is the detuning of the cavity, $g$ is the parametric coupling strength, $\theta_p$ is the relative phase between the signal and second harmonic waves. In our experiment, the vacuum channel of OPO2 is replaced by a squeezed vacuum, then the basis supplied by OPO2 is changed by a squeezing unitary transformation. Hence, we should substitute the operators as $\hat{a} = \hat{a}_s \cosh r - \hat{a}_s^+ e^{-i\theta} \sinh r$, $\hat{a}^+ = \hat{a}_s^+ \cosh r - \hat{a}_s e^{i\theta} \sinh r$, where $\theta$ is the relative phase between the pump and signal waves of OPO1, and $r$ is the squeezing parameter. In the new squeezed basis, the Hamiltonian is given as

$$\begin{aligned}\hat{H}_s &= \Delta_c \hat{a}^+ \hat{a} + \frac{1}{2}\hbar g e^{i\theta_p} \hat{a}^2 + \frac{1}{2}\hbar g e^{-i\theta_p} (\hat{a}^+)^2 \\ &= (\hat{a}_s)^2 e^{i\theta} \left[ \frac{1}{2}\hbar g \left( e^{i(\theta_p - \theta)} \cosh^2 r + e^{-i(\theta_p - \theta)} \sinh^2 r \right) - \Delta_c \cosh r \sinh r \right] \\ &\quad + (\hat{a}_s^+)^2 e^{-i\theta} \left[ \frac{1}{2}\hbar g \left( e^{i(\theta_p - \theta)} \sinh^2 r + e^{-i(\theta_p - \theta)} \cosh^2 r \right) - \Delta_c \cosh r \sinh r \right] \\ &\quad + \hat{a}_s^+ \hat{a}_s \left[ \Delta_c (\cosh^2 r + \sinh^2 r) - \hbar g \cosh r \sinh r \left( e^{i(\theta_p - \theta)} + e^{-i(\theta_p - \theta)} \right) \right] \\ &\quad + \left[ \Delta_c \sinh^2 r - \frac{1}{2}\hbar g \cosh r \sinh r \left( e^{i(\theta_p - \theta)} + e^{-i(\theta_p - \theta)} \right) \right]\end{aligned} \quad (S2.2)$$

Choosing $\theta_p - \theta = \pi$, then it is simplified to

$$\begin{aligned}\hat{H}_s &= -(\hat{a}_s)^2 e^{i\theta} \left( \frac{1}{2}\hbar g \cosh 2r + \frac{1}{2}\Delta_c \sinh 2r \right) \\ &\quad - (\hat{a}_s^+)^2 e^{-i\theta} \left( \frac{1}{2}\hbar g \cosh 2r + \frac{1}{2}\Delta_c \sinh 2r \right) \\ &\quad + \hat{a}_s^+ \hat{a}_s (\Delta_c \cosh 2r + \hbar g \sinh 2r) + \left( \Delta_c \sinh^2 r + \frac{1}{2}\hbar g \sinh 2r \right) \\ &= -\frac{1}{2}\hbar g \cosh 2r \left[ (\hat{a}_s)^2 e^{i\theta} + (\hat{a}_s^+)^2 e^{-i\theta} \right] \\ &\quad - \frac{1}{2}\Delta_c \sinh 2r \left[ (\hat{a}_s)^2 e^{i\theta} + (\hat{a}_s^+)^2 e^{-i\theta} \right] \\ &\quad + \hat{a}_s^+ \hat{a}_s (\Delta_c \cosh 2r + \hbar g \sinh 2r) \\ &\quad + \left( \Delta_c \sinh^2 r + \frac{1}{2}\hbar g \sinh 2r \right)\end{aligned} \quad (S2.3)$$

In the Heisenberg picture, the time evolution of an operator follows the Heisenberg equation. The equations of motion inside the OPO2 can be expressed as $\dot{\hat{a}}_s = \frac{1}{i\hbar}[\hat{a}_s, H_s] - \frac{\kappa}{2}\hat{a}_s + \sqrt{\kappa}\hat{b}_{in}$, $\hat{b}_{in}$ is the annihilation operator of the squeezed vacuum input [S5]. We substitute Eq. (S2.3) into the motion equation of OPO2, the resulting equation is written as

$$\frac{d}{dt}\hat{a}_s = \left( -\frac{i\Delta_c}{\hbar}\cosh 2r - ig \sinh 2r \right)\hat{a}_s + \left( \frac{i\Delta_c}{\hbar}\sinh 2r + ig \cosh 2r \right)e^{-i\theta}\hat{a}_s^+ - \frac{\kappa}{2}\hat{a}_s + \sqrt{\kappa}\hat{b}_{in}$$

(S2.4)

$$\frac{d}{dt}\hat{a}_s^+ = \left(-\frac{i\Delta_c}{\hbar}\sinh 2r - ig\cosh 2r\right)e^{i\theta}\hat{a}_s + \left(\frac{i\Delta_c}{\hbar}\cosh 2r + ig\sinh 2r\right)\hat{a}_s^+ - \frac{\kappa}{2}\hat{a}_s^+ + \sqrt{\kappa}\hat{b}_{in}^+$$

(S2.5)

We define $A = -\frac{i\Delta_c}{\hbar}\cosh 2r - ig\sinh 2r$, $B = \left(\frac{i\Delta_c}{\hbar}\sinh 2r + ig\cosh 2r\right)e^{-i\theta}$, then the equations can be rewritten by taking the Fourier transform in matrix form

$$\begin{pmatrix} -i\omega & 0 \\ 0 & i\omega \end{pmatrix}\begin{pmatrix} \hat{a}_s \\ \hat{a}_s^+ \end{pmatrix} = \begin{pmatrix} A - \frac{\kappa}{2} & B \\ B^* & A^* - \frac{\kappa}{2} \end{pmatrix}\begin{pmatrix} \hat{a}_s \\ \hat{a}_s^+ \end{pmatrix} + \sqrt{\kappa}\begin{pmatrix} \hat{b}_{in} \\ \hat{b}_{in}^+ \end{pmatrix}$$

(S2.6)

The solutions of annihilation operator $\hat{a}_s$ and creation operator $\hat{a}_s^+$ of the intracavity mode can be solved as

$$\hat{a}_s = \frac{\sqrt{\kappa}}{(-i\omega - A + \frac{\kappa}{2})(i\omega - A^* + \frac{\kappa}{2}) - |B|^2}\left[\left(i\omega - A^* + \frac{\kappa}{2}\right)\hat{b}_{in} + B\hat{b}_{in}^+\right]$$

(S2.7)

$$\hat{a}_s^+ = \frac{\sqrt{\kappa}}{(-i\omega - A + \frac{\kappa}{2})(i\omega - A^* + \frac{\kappa}{2}) - |B|^2}\left[B^*\hat{b}_{in} + \left(-i\omega - A + \frac{\kappa}{2}\right)\hat{b}_{in}^+\right]$$

(S2.8)

Evidently, $(\hat{a}_s)^+ = \hat{a}_s^+$, they are Hermitian conjugate with each other.

Based on the beam splitter model [S6], $\hat{a}_{out} = \sqrt{\kappa}\hat{a}_s + \sqrt{1-\kappa}\hat{b}_{in}$ and $\hat{a}^+{}_{out} = \sqrt{\kappa}\hat{a}_s^+ + \sqrt{1-\kappa}\hat{b}_{in}{}^+$, the photon number operators of the signal wave exiting from OPO2 under the cavity boundary conditions are deduced to

$$\hat{a}_{out} = \sqrt{\kappa}\frac{\sqrt{\kappa}}{(i\omega - A + \frac{\kappa}{2})(-i\omega - A^* + \frac{\kappa}{2}) - |B|^2}\left[\left(-i\omega - A^* + \frac{\kappa}{2}\right)\hat{b}_{in} + B\hat{b}_{in}^+\right] + \sqrt{1-\kappa}\hat{b}_{in}$$

(S2.9)

$$\hat{a}_{out}^+ = \sqrt{\kappa}\frac{\sqrt{\kappa}}{(i\omega - A + \frac{\kappa}{2})(-i\omega - A^* + \frac{\kappa}{2}) - |B|^2}\left[B^*\hat{b}_{in} + \left(i\omega - A + \frac{\kappa}{2}\right)\hat{b}_{in}^+\right] + \sqrt{1-\kappa}\hat{b}_{in}^+$$

(S2.10)

where $\hat{b}_{in}$ represents the annihilation operator of the injected squeezed vacuum. $\kappa$ is the cavity damping constant. From Eq. (S2.9) and Eq. (S2.10), we can also find that $(\hat{a}_{out})^+ = \hat{a}_{out}^+$, which are Hermitian conjugate with each other. From the above equations, we can calculate the mean photon number as $\hat{n}_{out} = \hat{a}_{out}{}^{+}\hat{a}_{out}$, which is a linear dependence with the pump power.

The *X*-quadrature and *P*-quadrature of the signal wave are written as $\hat{x} = \frac{1}{2}(\hat{a} + \hat{a}^{+})$ and $\hat{p} = \frac{1}{2i}(\hat{a} - \hat{a}^{+})$, and we can rewrite $\hat{b}_{in} = \hat{x}_{in} + i\hat{p}_{in}$, respectively. Then, the *X*-quadrature of signal wave becomes

$$\hat{x}_{out} = \left( \frac{\kappa^2}{2(i\omega - A + \sqrt{\kappa})(-i\omega - A^* + \sqrt{\kappa})} + \sqrt{1-\kappa} \right) \hat{x}_{in}$$
$$+ i \left[ \frac{\kappa}{(i\omega - A + \sqrt{\kappa})(-i\omega - A^* + \sqrt{\kappa})}(-i\omega + A - B) \right] \hat{p}_{in} \quad \text{(S2.11)}$$

We obtain the noise variance of *X*-quadrature as

$$\delta^2 \hat{x}_{out} = \left( \frac{\kappa^2}{2(i\omega - A + \sqrt{\kappa})(-i\omega - A^* + \sqrt{\kappa})} + \sqrt{1-\kappa} \right)^2 \delta^2 \hat{x}_{in} -$$
$$\left[ \frac{\kappa}{(i\omega - A + \sqrt{\kappa})(-i\omega - A^* + \sqrt{\kappa})}(-i\omega + A - B) \right]^2 \delta^2 \hat{p}_{in} \quad \text{(S2.12)}$$

Substituting the quadrature noise variances $\delta^2 \hat{x}_{in} = e^{2r}$ and $\delta^2 \hat{p}_{in} = e^{-2r}$ of a phase squeezed state into Eq. (S2.12), the noise variance of *X*-quadrature is

$$\delta^2 \hat{x}_{out} = \left( \frac{\kappa^2}{2(i\omega - A + \sqrt{\kappa})(-i\omega - A^* + \sqrt{\kappa})} + \sqrt{1-\kappa} \right)^2 e^{2r} -$$
$$\left[ \frac{\kappa}{(i\omega - A + \sqrt{\kappa})(-i\omega - A^* + \sqrt{\kappa})}(-i\omega + A - B) \right]^2 e^{-2r} \quad \text{(S2.13)}$$

It can be found that the noise variance of *X*-quadrature not only depends on the strength of parametric interaction *g* in OPO2, but also relates to the injected squeezing parameter *r*.

Based on the master equation theory [S7], the enhanced parametric interaction process is usually accompanied by an undesired noise generation emerging from the two-photon damping effect, which can be equivalent to vacuum noise coupling from a beam splitter. Taking the actual experimental situation into consideration, we cannot neglect the influence of two-photon damping effects. Therefore, we introduce an additional noise term by the employment of the beam splitter model to amend the result of Eq. (S2.13) to reflect the contribution of the two-photon damping effect [S6, S8, S9]. Here, the noise variance is measured at the sideband frequency of 18 MHz, at which the laser intensity noise reaches shot noise limit. After taking the two-photon damping effect into consideration, the final quadrature noise variance of the output of OPO2 can be rewritten as [S6, S8, S9]

$$\delta^2 \hat{x}'_{out} = \eta \delta^2 \hat{x}_{out} + 1 - \eta \tag{S2.14}$$

where $1-\eta$ is the efficient optical loss relating to two-photon damping, which associates to the two squeezing parameters, i.e., is defined as $\eta = 1 - \alpha * \Delta r$. $\Delta r = |r - r_P|$, $r_P$ is the squeezing parameter of OPO2, and $\alpha$ is a constant coefficient. We can fit the experimental data of Fig. 4(b) with Eqs. (S2.13) and (S2.14), and obtain $\alpha = 2.1$ with $r_P = 1.15$. As results, the experimental data (dots) are on good agreement with the theoretical analysis of Eq. (S2.14), rather than Eq. (S2.13), which verify that the model is correct.

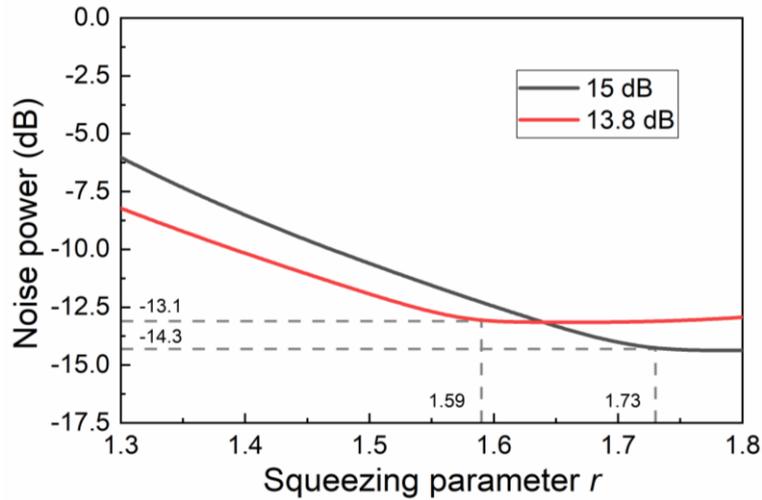

Fig. S2 Noise powers of the squeezed laser versus the squeezing parameters of the reservoir, based on the state-of-the-art OPOs with quantum noise reduction of 13.8 dB [S10] and 15 dB [S11], respectively.

Based on the theoretical models above, we simulate the noise powers of the squeezed laser based on the state-of-the-art squeezed vacuum states in references [S10, S11]. We suppose that the noise power of the squeezed basis is 13.8 dB (1.59) or 15 dB (1.73) below the shot noise limit. In the simulation, the experimental parameters are given as $\kappa = 15\%$, $g = \frac{P^2}{2\omega\hbar} * d_{eff} * \omega_0 * L$, $\omega = 18*10^6$ Hz, $d_{eff} = 10*10^{-12}$ m/V, $\Delta_c = 2.8*10^3$ Hz, $\omega_0 = 41$ μm, $L=10$ mm. Then, the noise power of the squeezed laser is calculated relating to the squeezing parameter $r$ of the squeezed vacuum reservoir as shown in Fig. S2. It can be seen that the final squeezing level of the squeezed laser is limited by the squeezing parameter of the squeezed basis, and stronger squeezing injection eliminates more undesired noise to generate higher squeezing degree of squeezed laser. With the state-of-the-art OPO, a -14.3 dB squeezed laser can be generated without considering the propagation losses between the two OPOs.

S**3. Laser property of the squeezed lasing**

(i) **Single mode operation with frequency-degenerate OPO.**

An OPO intrinsically operates at multi-cavity modes with a frequency interval of the cavity's free spectral range (FSR) [S12, S13]. Here, we employ two OPOs to generate squeezed lasing. OPO1 is employed for squeezed vacuum preparation, and OPO2 is applied for squeezed basis supply and parametric amplification. By constructing two OPOs with different cavity length, we ensure a stable single-cavity-mode operation of the OPO2 under squeezing enhanced regime. The physical mechanism is explained in detail by the analysis below.

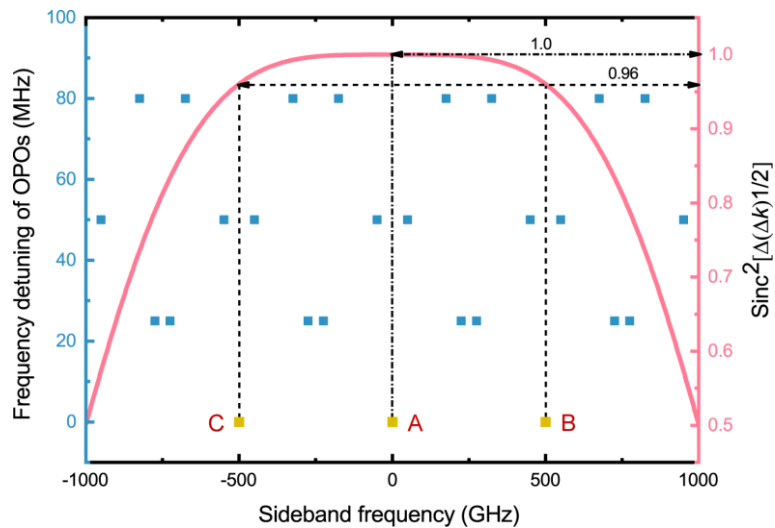

Fig. S3 Frequency detuning of the two OPOs in the nonlinear acceptance bandwidth. Blue points indicate the frequency detuning between OPO1 and OPO2, the longitudinal axis is the frequency

detuning. A frequency detuning means that the corresponding sideband mode cannot oscillate above the threshold of the OPO2 under our pump condition, owing to the lack of corresponding squeezed sideband modes generated from the OPO1.

The output of the OPO has a frequency comb distribution characteristic divided by the FSR of the cavity, covering the nonlinear spectral acceptance bandwidth of the PPKTP [S14]. The normalized transmission intensity of the frequency comb sidebands of an optical cavity is expressed as [S12, S13]

$$t_{OPO} = \left| \frac{\sqrt{(1-r_1^2)(1-r_2^2)}e^{\frac{i\pi\upsilon}{FSR}}}{1-r_1 r_2 e^{\frac{2i\pi\upsilon}{FSR}}} \right|^2 \tag{S3.1}$$

where $r_1$ and $r_2$ represent the reflectivity of the two cavity mirrors. $\upsilon$ is the sideband frequency [S13]. The FSR is expressed as $\frac{c}{2L_{1,2}}$, in which $L_{1,2}$ is the cavity length of the two OPOs. In the experiment, the spectral acceptance bandwidth of a 10 mm PPKTP crystal at 1064 nm is 2 THz. The two OPOs have the length of $L_1$=37 mm and $L_2$=31 mm, corresponding to the FSRs of $FSR_1$=3.33 GHz and $FSR_2$=3.76 GHz.

Within the spectral acceptance bandwidth, we calculate the frequency comb sidebands of the two OPOs with Eq. (S3.1). The frequency detuning of OPO1 with respect to OPO2 is illustrated in Fig. S3. Supposed that the resonance frequencies of the two OPOs overlap at the central point of the spectral acceptance bandwidth, these neighboring longitudinal modes do not simultaneously resonant at two OPOs, but have a frequency detuning (approximately 25 MHz, 50 MHz, 80 MHz). Therefore, these neighboring longitudinal modes cannot oscillate above the threshold of the OPO2 under our pump condition, owing to the lack of corresponding squeezed sideband modes generated from the OPO1. Until the sideband frequency reaches ±500 GHz that is lowest common multiple of the FSRs of two OPOs, corresponding to points B and C, and the sideband modes of two OPOs become co-resonant. Fig. S3 shows that only three sideband modes in the spectral bandwidth can resonate simultaneously. Fortunately, the sideband mode at the point A has higher nonlinear conversion efficiency than that of point B and C. Therefore, the sideband mode at point A starts to oscillate prior to that of point B and C. Under the condition of low pump ratio, the single-cavity-mode operation can be kept.

Here, the injected squeezing is -8.6 dB, we observed a stable single cavity mode operation of OPO2 at the pump ratio from 1 to 1.75. The single longitudinal mode operation is experimentally confirmed by the noise spectrum output from SHD or optical spectrum analyzer. With the pump ratio going beyond 1.75, multi-cavity-mode begins to oscillate. Therefore, we set the pump ratio below 1.75 in the experimental preparation of squeezed laser to ensure single-cavity-mode operation. The phenomena can be explained by a mode-competition mechanism.

**(ii) Stimulated parametric coupling induced by exponentially enhanced gain**

Beginning with a subthreshold OPO used as an amplifier, spontaneous parametric coupling dominates the interaction in the cavity, which has a parametric gain of $\sim e^{r_p}$ [S15, S16]. However, with sufficient squeezed vacuum reservoir at opposite phase and with constant pump power, the OPO can be transferred into superthreshold performance, in which the parametric gain is enhanced to $\sim \cosh 2r$ [S7]. By comparing the gain factors under the two cases as shown in Fig. S4, we can see that the enhanced gains are nearly indistinguishable as the squeezing parameter below 0.8. However, with higher level squeezing input, the second case can obtain greater amplification to produce a stimulated parametric coupling. Therefore, a lasing is generated by only controlling the injected squeezing parameter.

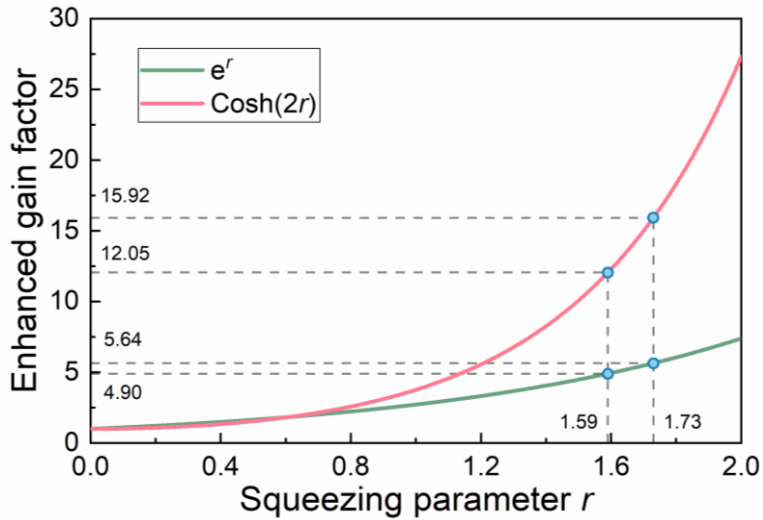

Fig. S4 The relationship between the gain amplification factor and the injection squeezing

parametric *r* under the two mechanisms of spontaneous emission and stimulated emission.

Furthermore, in the experiment, the squeezed vacuum reservoir has a bright amplitude, which is also injected into OPO2. But we have experimentally demonstrated that the weak mean field is not amplified in the squeezed lasing process.

Firstly, we have observed the squeezed laser performance under squeezed vacuum injection without the weak mean field, i.e., no seed beam is injected into OPO1. By comparing the squeezed laser powers with and without the seed beam, we have found that the squeezed laser powers exporting from OPO2 are basically the same for the two cases (~2.6 mW). It is inferred that the weak mean field is not associated to the observed change in the power in Fig. 3b.

Table. S3 Results of the output power of the phase squeezed light exporting from OPO1

| Pump Power (mW) | 0 | 30 | 35 | 40 | 45 | 50 | 55 | 60 | 65 | 70 |
|---|---|---|---|---|---|---|---|---|---|---|
| Output Power (μW) | ~3 | 19.5 | 20.44 | 23.1 | 29.91 | 33.8 | 47.4 | 63 | 85.4 | 112.6 |

Secondly, under a seed beam injecting, we have measured the powers of the weak mean field exporting from OPO1 by changing its pump power from 30 mW to 70 mW, which increases from 19.5 μW to 112.6 μW (Table. S3). By comparing the optical powers of the weak mean field with the squeezed laser in Fig. 3b, we can indirectly conclude that the weak mean field has little contribution to the laser power. We give a detailed analysis as follows.

For example, we assume that the squeezed laser power is completely attributed by the amplification of the weak mean field. Here, we list the weakest and strongest weak mean field exporting from OPO1 as a contrast, which is shown in Table S4. By comparing the amplification factor of the two cases, it is decreased from 98 to 24.3. In our experiment, more weak mean field means stronger squeezing injection. That is to say, the parametric gain in OPO2 is reduced as the injected squeezing level increased. Apparently, it is contradictory with the exponentially enhanced gain regime in our system, i.e., $g_s = g\cosh 2r$. Hence, the assumption is invalid.

Table. S4 Comparation of the remaining injected power and squeezed laser power of OPO2.

| Total loss from OPO1 to OPO2 | The remaining injected power (μW) | | Squeezed laser power (μW) | Amplification factor |
|---|---|---|---|---|
| 5% | Min | 18.5 | 1.82 | 98 |
| | Max | 107 | 2.6 | 24.3 |

Additionally, as stated by the Ref. [S7], it is true that the squeezed vacuum carried by the weak mean field is applied to eliminate the undesired noise, and the squeezing enhanced amplification only acts on the interaction corresponding to the squeezed basis. By combing the above analysis, we can conclude that the weak mean field is not associated to the observed change in the power in Fig. 3b. And only the photons generated by OPO2 is amplified due to a phase sensitive amplification process in the cavity. The main cause of this phenomenon is attributed to the opposite phase of the weak mean field to the squeezed laser field, which is deamplified by a destructive interference process. Therefore, only the photons generated in OPO2 can be exponentially amplified to contribute to the power of the squeezed laser.

**(iii) Dependence of threshold power on injected squeezing parameter $r$**

In general, the threshold power of a doubly-resonant OPO can be expressed as [S17]

$$P_{th} = \frac{(T_P + V_P)^2 (T_s + l)^2}{2 T_P E} \tag{S3.2}$$

where, $E$ is the nonlinear coefficient of the parametric interaction in the cavity, and is written as [S18]

$$E = \frac{4\mu_0 d_{eff}^2 \omega_s^2 \omega_i^2 L h(\xi)}{\pi c^2 n_P^2 \omega_P} \tag{S3.3}$$

where, the Gaussian beam focusing factor is $h(\xi) \approx 1$, the vacuum permeability is $\mu_0 = 4\pi*10^{-7}$, and $n_P = 1.89$ is the refractive index of the pump field. $\omega_{s,i} = 1.77*10^{15}$ and $\omega_P = 3.54*10^{15}$ are the angular frequencies of the signal or idle and pump fields, respectively. The transmittance of the pump and signal fields are $T_P = 0.025$ and $T_S = 0.15$, respectively. $V_P = 0.011$ and $l = 0.004$ are the round-trip losses of the pump and signal field in the cavity. With these parameters, the theoretical threshold power of OPO2 is predicted to be ~45 mW, which is in accordance with the measured value of the OPO.

Additionally, the classical parametric gain $g_0$ of the OPO relating to the threshold

power can be expressed as [S6]

$$g_0 = \frac{1}{(1-\sqrt{P/P_{th}})^2} \quad (S3.4)$$

In our reservoir-engineered OPO system, the parametric gain is exponentially enhanced by a factor of cosh2$r$. Therefore, we can foresee that the threshold power of the OPO is reduced by the enhanced gain factor of cosh2$r$. We do an experiment to calibrate the threshold powers under different squeezing parameters as shown in Fig. S5. It can be seen that the threshold power of OPO2 is dramatically decreased with the squeezing parameter being increased. For example, when the squeezing parameter changes from 0.8 to 0.99, the threshold power is reduced from 37 mW to 11.4 mW. This effect mainly relates to the squeezed noise level in the cavity, which is raised by stronger squeezing injection to remove more undesired noise as shown in Fig. S2. Therefore, the raised squeezing level bring into larger amplification times of the intracavity photons, which further decreases the threshold power of OPO2. However, the dynamical process between the threshold power and the squeezing and anti-squeezing noise variances is extremely complicated, where we only show the experimental results to support our point.

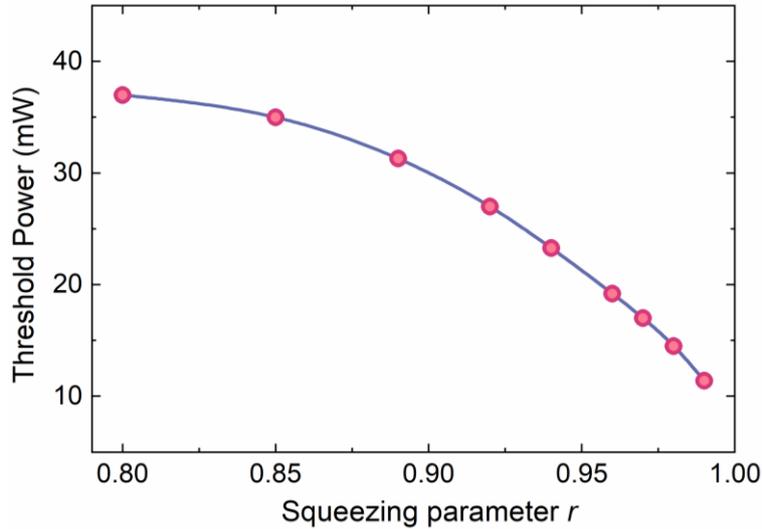

Fig. S5 OPO2 threshold power versus the injected squeezing parameter $r$.

**(iv) Laser linewidth characterization.**

The last property of the squeezed laser is the coherence preservation, which is demonstrated by the characterization of laser linewidth. Here, we measure the linewidth with a β-separation line method by integrating the frequency noise power spectral density (PSD) [S19, S20]. The laser linewidth based on the PSD can be read as [S20]

$$\Gamma = (8\ln(2)D)^{1/2} \tag{S3.5}$$

where $D$ represents the surface of the high modulation index region, which is integrated with the overall surface under the portions of $S_{\delta\nu}(f)$ exceeding the β-separation line. More details can be found in Ref. [10]. $D$ and $S_{\delta\nu}(f)$ are expressed as

$$D = \int_{1/T_0}^{\infty} H(S_{\delta\nu}(f) - 8\ln(2)f/\pi^2) S_{\delta\nu}(f) df \tag{S3.6}$$

$$S_{\delta\nu}(f) = \frac{10^{(PSD/10)}/1000}{k_0^2/Z_0}/RBW \tag{S3.7}$$

where $H(x)$ is the Heaviside unit step function. $H(x)$=1, if $x \geq 0$; $H(x)$=0, if $x < 0$).

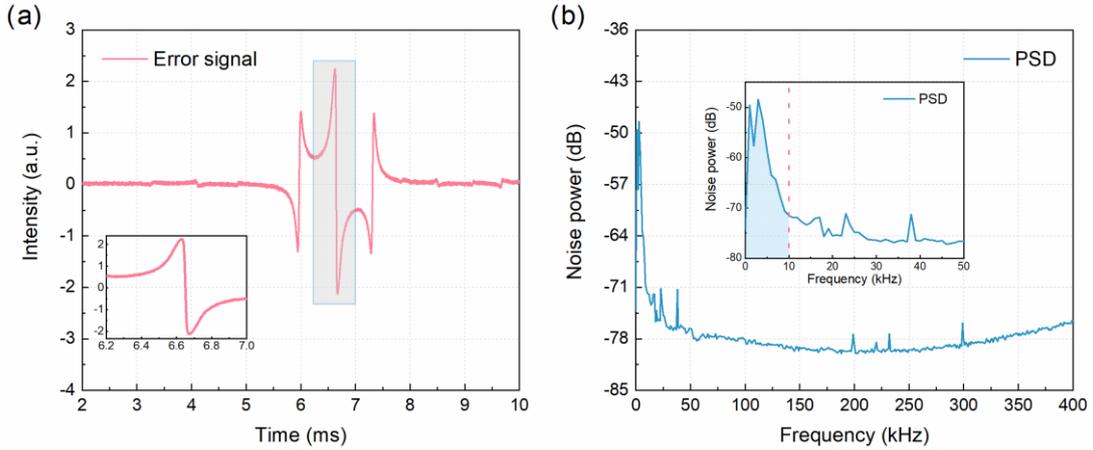

Fig. S6 (a) Error signal of MC3. (b) PSD of the squeezed laser reflected from MC3, RBW: 100 Hz and VBW: 10 Hz.

In our experiment, the PDH error signal of a mode cleaner (MC) is used for frequency noise $S_{\delta\nu}(f)$ measurement [S21]. In this case, $k_0$ represents the slope of the PDH error signal in MC3's length locking loop, which is determined by scanning the cavity length of MC3 and reading out on the oscilloscope, as shown in Fig. S6(a). The impedance of the SA is $Z_0$=50Ω. Then, the error signal is divided into two parts. One is feedback to the PZT sticked on the cavity mirror to lock the length of MC3. The other one is utilized to extract the frequency noise PSD, which is shown in Fig. S6(b). Then, we can calculate the laser linewidth with Eqs. (S3.5) - (S3.7). Here, we measure the

PSDs of the 1064 nm laser source and squeezed laser with MC1 and MC3, respectively. We take one of the PSDs of squeezed lasing as an example, i.e., $r$=0.99, which is shown in Fig. S6(b). It is apparent that the intersection point between $\beta$-separation line and the PSD is around 10 kHz. Finally, the linewidth $\Gamma$ of the squeezed laser can be read out by the shadow area in Fig. S6(b), which is 15 kHz. The measured linewidth equals to the linewidth of 1064 nm laser source. The other linewidths with different squeezing parameters are listed in Table. S5 and also shown in Fig. 3(c) of the main text. The results demonstrate a coherence characteristic preservation of the squeezed laser.

Table. S5 Results of the measured linewidths of the 1064 nm and squeezed lasers

|  | Squeezed laser | | | | | 1064nm laser |
|---|---|---|---|---|---|---|
| **Squeezing parameter $r$** | 0.80 | 0.89 | 0.94 | 0.97 | 0.99 |  |
| **Laser linewidth $\Gamma$ (kHz)** | 30 | 27.8 | 23 | 18.7 | 15 | 15 |